\documentclass[twocolumn,aps,prl,superscriptaddress,showpacs,floatfix]{revtex4}
\usepackage{amsmath,bm}
\usepackage{graphicx}
\usepackage{epsfig}
\begin{document}

\title{Entropy Production at RHIC}
\author{Subrata Pal}
\affiliation{National Superconducting Cyclotron Laboratory
and Department of Physics and Astronomy, Michigan State University, East
Lansing, Michigan 48824}
\author{Scott Pratt}
\affiliation{Department of Physics and Astronomy, Michigan State
University, East Lansing, Michigan 48824}

\begin{abstract}
For central heavy ion collisions at the RHIC energy, the entropy per unit
rapidity $dS/dy$ at freeze-out is extracted with minimal model dependence from
available experimental measurements of particle yields, spectra, and source
sizes estimated from two-particle interferometry. The extracted entropy rapidity
density is consistent with lattice gauge theory results for a thermalized 
quark-gluon plasma with an energy density estimated from transverse energy 
production at RHIC.

\end{abstract}
\maketitle

Lattice quantum chromodynamics (QCD) calculations indicate \cite{lattice} that
at zero net baryon density at a critical temperature of $T_c \sim 170$ MeV and
energy density $\epsilon_c \sim 1$ GeV/fm$^3$ a color deconfined and a chirally
restored quark-gluon plasma (QGP) phase is formed. These energy densities are
attained in relativistic heavy ion collisions, where it is believed that a
dense system of quarks and gluons is created which undergoes rapid collective
expansion before the partons hadronize and eventually decouple.

The ongoing experimental program at RHIC is devoted to the search for the
signals of QGP. Unfortunately, measurements are confined to final-state
particles which are mostly hadrons.  One interesting quantity that may provide
valuable insight into the state of the matter in the early stages of the
collision is the net entropy which is roughly conserved between initial
thermalization and freeze-out \cite{Siemens,Letes}. After freeze-out, when
particles are free-streaming, the entropy is fixed. Entropy would also be
conserved during the expansion stage if the mean free paths were very
short. Viscous effects, shock waves, and the decoupling process can only
increase the entropy. Thus, the entropy from the final state provides an upper
bound for the entropy of the initial state.

Compared to a pion gas at the same energy density, the QGP is a high-entropy
state as can be understood by considering the equation,
\begin{equation}
\label{eq:DeltaS}
\Delta S=\frac{Q}{T},
\end{equation}
where $\Delta S$ is the change in entropy resulting from the addition of an
amount of heat $Q$ into a fixed volume at temperature $T$. Lattice calculations
show that the transition is nearly a first order with a distinct peak in the
specific heat near $T_c$. In this region, the temperature stays relatively
fixed while the energy density changes substantially. Thus, the existence of a
latent heat keeps the temperature lower than a pion gas at the same energy
density, which results in a higher entropy as can be seen from
Eq. (\ref{eq:DeltaS}). At high energy densities, the temperature of a QGP
remains lower than that of the pion gas due to the disparity in the number of
degrees of freedom. Asymptotically, the energy density of a QGP approaches that
of a massless gas of quarks and hadrons,
\begin{equation}
\epsilon=\left(N_B+\frac{7}{8}N_F\right)\frac{\pi^2}{30}T^4,
\end{equation}
where $N_B$ and $N_F$ are the number of Bosonic and Fermionic degrees of
freedom. Whereas a pion gas has 3 bosonic degrees of freedom, a QGP has 47.5
effective degrees of freedom if one considers gluons and the three light quark
flavors. Thus, relative to a pion gas, the QGP has a lower temperature and
higher entropy. If the partonic degrees of freedom were not fully liberated at
RHIC, one would thus expect a lower entropy for the same energy
density. Furthermore, if one were to consider a full hadron gas, incorporating
all the resonances from the particle data book at energy densities far above
the point where the hadronic models are justified, one would expect to have a
state of matter with even more entropy than a QGP for temperatures near or
above 250 MeV.

The entropy density per unity rapidity $dS/dy$ can be calculated from the final
phase space density, $f({\bf p},{\bf r},t_f)$, evaluated at the freeze-out time
$t_f$. Spectra provide information only about the final momenta, ${\bf p}$. The
spatial information in $f$ must be determined by correlations which can provide
the characteristic dimensions of $f({\bf p},{\bf r},t)$ for given values of
${\bf p}$ \cite{Pratt,Bertsch2}. In fact, it has been suggested that the softer
equation of state associated with a first order phase transition to QGP with a
large latent heat results in long lived and/or large source sizes
\cite{Bertsch,Pratt2,Rischke,Soff,teaneyhydro}. The fact that the QGP models
tend to accurately reproduce spectra while over-predicting source sizes
\cite{Soff,teaneyhydro,heinzkolb,kolbheinz,kolbheinz2} suggests
that the entropy at RHIC might fall below the expectations for a quark-gluon
plasma.

In this letter we extract $dS/dy$ for hadrons at freeze-out from yields and
spectra combined with the source dimensions in central Au+Au collisions at the
RHIC energy of $\sqrt{s} = 130A$ GeV. By comparing this estimated net hadron
$dS/dy$ with the entropy at the early stage in the parton medium, we address
several key issues concerning the initial stage of the collisions and the
equilibration of partonic degrees of freedom.

In the Landau quasi-particle approximation the entropy is given by
\begin{eqnarray}\label{entropy}
S &=& (2J+1)\int \frac{d^3r d^3p}{(2\pi)^3} \left[ -f\ln f \pm
(1\pm f)\ln (1\pm f) \right] \nonumber \\ 
&\approx& (2J+1)\int \frac{d^3r d^3p}{(2\pi)^3} \left[ -f\ln f + f \pm f^2/2\right],
\end{eqnarray}
where the $\pm$ sign corresponds to bosons/fermions with total spin $J$.  In
the classical limit $f\ll 1$, the $f^2$ and higher orders of $f$ can be
neglected. The classical limit is sufficient for all particles except for
pions. Even for pions, the $f^2$ term contributes only on the 3\%
level. Assuming a three-dimensional Gaussian form for the density
profile in coordinate space, the single-particle phase space density 
at any time $t$ at or after freeze-out \cite{Bertsch,Tomasik} is
\begin{equation}\label{phase}
f({\bf r},{\bf p},t) = f_{\rm max}({\bf p})  \exp\left(
-\frac{x^2}{2R_x^2} - \frac{y^2}{2R_y^2} - \frac{z^2}{2R_z^2} \right),
\end{equation}
where $R_x$, $R_y$ and $R_z$ are implicit functions of ${\bf p}$. Since spectra
are obtained from integrating the phase space density, $f_{\rm max}$, and
therefore $f({\bf p},{\bf r},t)$, can be determined from the radii and spectra.
\begin{eqnarray}\label{momd}
\frac{dN}{d^3p} 
&=& (2J+1)\int \frac{d^3r}{(2\pi)^3} f({\bf r},{\bf p}) ~, \nonumber \\
f_{\rm max}({\bf p}) 
&=& \frac{(2\pi)^{3/2}}{(2J+1)}\frac{dN}{d^3p} \frac{1}{R^3} ~,
\end{eqnarray}
where $R^3$ is the product of the three radii. Chemical equilibration has not
been invoked as the expressions do not involve temperature or chemical
potentials. Instead the phase space densities and therefore the entropies will
be functions of the spectra and the source sizes which can be inferred from
experiment.

In Hanbury-Brown Twiss (HBT) correlation measurements for identical particles, 
if the single-particle
emission source is characterized by a Gaussian as in Eq. (\ref{phase}), the
Gaussian widths are each inversely proportional to the source dimensions in the
canonically conjugate spatial variables \cite{Pratt3}. One then obtains the
two-particle correlation function in terms of the radii,
\begin{eqnarray}\label{corrf}
C({\bf K},{\bf q}) &=& \frac{P({\bf p}_1, {\bf p}_2)}{P({\bf p}_1) P({\bf p}_2)} \nonumber\\
&=& 1 + \exp\left( - q_o^2R_o^2 - q_s^2R_s^2 - q_l^2R_l^2 \right) , 
\end{eqnarray}
where ${\bf K}=({\bf p}_1+{\bf p}_2)/2$ and ${\bf q} = {\bf p}_1 - {\bf p}_2$.
The momentum difference $q$ has been decomposed into ``out-side-long"
components defined by the beam direction $q_l$, parallel to the total momentum
of the pair in the transverse plane $q_o$, and the direction orthogonal to the
above two direction $q_s$ \cite{Bertsch,Pratt2}. Data are generally analyzed in
the longitudinal co-moving system (LCMS) frame in which the longitudinal
component of the pair momentum vanishes. 

The Gaussian widths of Eq. (\ref{corrf}) in the LCMS frame are related to the
radii in the two-particle rest frame via a contraction factor $\gamma = E_T/m$. 
Phase space densities depend on the 
product of the three radii. In the rest frame of the pair, this product is
referred to as $R_{\rm inv}^3=\gamma R_oR_sR_l$.  Substituting the HBT
radii in Eq. (\ref{momd}), the entropy rapidity density from Eq.
(\ref{entropy}) in the classical limit is then
\begin{eqnarray}\label{eq:dsdy}
\frac{dS}{dy} &=& \int d^2p_T \: E\frac{dN}{d^3p} \left[ \frac{5}{2} 
- \ln {\cal F}({\bf p}) \pm \frac{1}{2^{5/2}} {\cal F}({\bf p}) \right] , \nonumber \\
{\cal F}({\bf p}) &=& \frac{1}{m} \frac{(2\pi)^{3/2}}{(2J+1)} \: E\frac{dN}{d^3p} 
\frac{1}{R_{\rm inv}^3} ~ .
\end{eqnarray}
Thus, $dS/dy$ of the final state can be estimated from the Gaussian source-size
parameters and spectra.

%%%%%%%%%%%%%%%%%%%%%%%%%%%%%%%%%%%%%%%%%%%%%%%
\begin{table*}
\caption{ The entropy content per unit rapidity, $dS/dy$, for hadrons produced 
at midrapidity in $\sim 11\%$ most central Au+Au collisions at the RHIC energy 
of $\sqrt s = 130A$ GeV.}

\begin{center}
\begin{tabular}{ccc|ccc|ccc|ccc|ccc|ccc|ccc|ccc} \hline\hline

\hfil & $\pi^-$ & \hfil & \hfil &  $\eta$ ($\eta'$) & \hfil &\hfil &  $K^-$ ($K^+$) 
& \hfil &\hfil & $\overline p$ ($p$) & \hfil &\hfil 
&  $\overline\Lambda \! + \! \overline\Sigma^0$ ($\Lambda \! + \! \Sigma^0$) 
& \hfil & \hfil & $\overline\Sigma^\pm$ ($\Sigma^\pm$) & \hfil & \hfil 
& $\overline\Xi^\pm$ ($\Xi^\pm$) & \hfil & \hfil & $\overline\Omega^+$ ($\Omega^-$) \\ \hline 
\hfil & 680 & \hfil & \hfil & 256 (69) & \hfil & \hfil & 256 (286) & \hfil & \hfil 
& 118 (159) & \hfil & \hfil & 81 (111) & \hfil & \hfil & 31 (42) & \hfil & \hfil 
& 23 (27) & \hfil & \hfil & 5 (5) \\ \hline\hline
 
\end{tabular}
\end{center}
\end{table*}
%%%%%%%%%%%%%%%%%%%%%%%%%%%%%%%%%%%%%%%%%%%%%%%%%%%%%%

The two-pion correlation function has been measured by STAR \cite{2pis} and
PHENIX \cite{2pip} for $\sim 11\%$ central Au+Au collisions at $\sqrt s = 130A$
GeV at midrapidity. In spite of about 70\% increase in pion multiplicity from
SPS to the RHIC energy the extracted radius, $R_{\rm inv} \approx 6.3$ fm at
RHIC, was very similar to what was measured at the SPS \cite{na44}. Moreover,
the prediction of $R_o$ larger than $R_s$ as a QGP signature was not
observed. Instead $R_o/R_s \approx 1$ was measured over a broad range of
transverse momenta $K_T=({\bf p}_{T_1} + {\bf p}_{T_2})/2$. The smaller values
of $R_l$ and $R_o$, as compared to hydrodynamic models, may be suggestive of a
rather high final phase space density
\cite{Bertsch2,Ray}, and therefore, a low entropy.
In our estimate of $\pi^-$ entropy we use the scaling relation $R_l =
A_l/\sqrt{m_T}$ where $m_T = \sqrt{K_T^2 + m_\pi^2}$ with the fitted value
$A_l=3.32\pm 0.03$ fm GeV$^{1/2}$ \cite{2pip}.  While for the $s$-$o$ direction
we employ the blast-wave inspired parameterization, $R_s^2 = R_o^2 = R_{\rm
geom}^2/(1 + \beta^2 m_T/T)$, where $R_{\rm geom} = 6.7 \pm 0.2$ fm, $\beta =
0.69$ and the freeze-out temperature is $T=125$ MeV
\cite{2pip}. The measured corresponding single particle $\pi^-$ spectra at RHIC
\cite{spectrap} (after correcting for electromagnetic and weak decays) is fitted 
to a Bose-Einstein distribution $EdN/d^3p = A'/[\exp(m_T/T_{\rm eff}) -1]$ 
where $A'=739$ GeV$^{-2}$ and effective (slope)
temperature $T_{\rm eff} = 197$ MeV. The entropy per unit rapidity estimated
from Eq. (\ref{eq:dsdy}) is $dS/dy|_{\pi^-} = 680$, and is shown in Table I.

Pions resulting from electromagnetic decays, i.e., those from $\eta$ and
$\eta'$ decays, should be subtracted from the spectra. From a thermal model with 
a chemical equilibrium temperature of 170 MeV, we estimate that 12\% of the pions come
from such decays and correct the spectra accordingly. We add in the entropy of
the $\eta$ and $\eta'$ by assuming that the $\eta$ contributes the same amount
of entropy as a single species of kaon which has a similar mass, and that the
$\eta'$ contributes half the amount of entropy as the average of protons and
anti-protons. The factor of 1/2 is applied to account for the lower spin
degeneracy of the $\eta'$. The entropies of the $\eta$ and $\eta'$ are
displayed in Table I. 

For a dilute gas in chemical equilibrium, it was demonstrated \cite{Siemens}
that the entropy per nucleon carried by the nucleons is related to the ratio of
the deuteron to the proton yield,
\begin{equation}
\label{eq:siemenskapusta}
 S = 5/2 -\ln{R_{dp}/3\sqrt 2} ~.
\end{equation}
If, within a given phase space cell, the ratio of a species $a$ to the product
of two species $b$ and $c$, which can undergo a reaction $a+b\leftrightarrow
c$, is determined by the ratio of available phase space for bound and unbound
states, one can determine the Gaussian source parameters from coalescence
arguments \cite{prattllope},
\begin{eqnarray}\label{pradi}
R^3_{\rm inv} &=& \pi^{3/2} \frac{(2J_c+1)}{(2J_a+1)(2J_b+1)} \nonumber \\ 
&& \times \left( \frac{E_a}{m_a} \frac{dN_a}{d^3p_a} \right) \!
\left( \frac{E_b}{m_b} \frac{dN_b}{d^3p_b} \right)
\!\! \Big/ \!
\left( \frac{E_c}{m_c} \frac{dN_c}{d^3p_c} \right) .
\label{eq:coalescence}
\end{eqnarray}
This expression is based on the assumption that the relative populations within
a given phase space cell are determined by the ratios of the available phase
space for bound and unbound states; whereas Eq. (\ref{eq:siemenskapusta})
requires the assumption that all phase space cells are equilibrated according
to a single chemical potential. Once $R_{\rm inv}$ is obtained as a function
of the momentum, one can extract the entropy from Eq. (\ref{eq:dsdy}).

\begin{figure}[ht]
\centerline{\epsfig{file=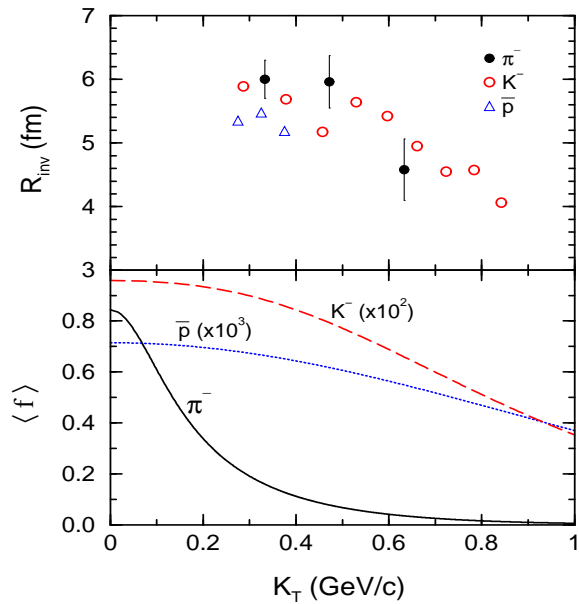,width=3.0in,height=3.2in,angle=0}}
\caption{ Top panel: The $K_T$ dependence of HBT radii for $K^-$ (open circles) 
and $\overline p$ (open triangles) estimated from invariant yields at midrapidity 
for central Au+Au collisions at $\sqrt s = 130A$ GeV. 
The measured $\pi^-$ radii (solid circles) from PHENIX \cite{2pip} are also shown.
Bottom panel: The average freeze-out phase space density as a function of 
$K_T$ for the hadrons.}

\label{fig:hbt}
\end{figure}

Assuming that $\overline p$ and $\overline n$ possess identical phase space densities,
which in turn is identical to that for $\overline d$ at the same relative velocity,
one can determine a source size for $\overline p$ using measured anti-proton and
anti-deuteron spectra. Figure \ref{fig:hbt} shows the $R_{\rm inv}$ for
antiprotons (open triangles) estimated (using Eq. (\ref{eq:coalescence})) from
the measured \cite{dbars} spectra of $\overline d$ and weak-decay-corrected 
$\overline p$s at three $K_T$ bins for the $\sim 18\%$ most central Au+Au collisions at
$\sqrt s = 130A$ GeV. As expected, the radii for the massive $\overline p$ exhibits
a smaller decrease with $K_T$ compared to that of pions. The $\overline p$
transverse momentum spectra for this centrality from the STAR \cite{pbars} and
the PHENIX \cite{spectrap} data are corrected for the weak hyperon decay
\cite{lambda} and matched in the overlap region. We fit this spectra in the
range $0.25<p_T<3.2$ GeV/c with a Gaussian distribution $EdN/d^3p = dN/dy
\exp(-p_T^2/2\sigma^2)/(2\pi\sigma^2)$ resulting in a yield of $dN/dy = 11.4$
and width $\sigma = 872$ MeV. Given the limited range and poor statistics for
anti-deuterons, we choose a constant average value, $R_{\rm inv}=5.34$ fm, independent
of $K_T$. Applying Eq. (\ref{eq:dsdy}), the entropy carried by the anti-protons
is then $dS/dy|_{\overline p} = 118$.

The procedure to estimate the $\overline p$ source radii can be identically applied
to extract the kaon radii using the $\phi$ meson, $\phi \leftrightarrow
K^+K^-$, in place of the deuteron. The coalescence arguments used in
Eq. (\ref{eq:coalescence}) ignore the binding energy of the composite particle,
which is justified if the binding energies are much less than the breakup
temperatures. This is an excellent assumption for the deuteron and a reasonable
assumption for the $\phi$ which is a resonance with a binding energy of $B=-30$
MeV. Although we neglect the binding energy in this analysis, one could adjust
the expression for $R_{\rm inv}^3$ by a factor of $e^{B/T}$ to account for the
binding energy \cite{prattllope}.

%%%%%%%%%%%%%%%%%%%%%%%%%%%%%%%%%%%%%%
%%%%%%%%%%%%%%%%%%%%%%%%%%%%%%%%%%%%%%

Figure \ref{fig:hbt} displays the $K^-$ invariant radii extracted from measured
midrapidity $\phi$ \cite{phis} and $K^-$ \cite{kaons} yields for the 11\% most
central Au+Au collisions at $\sqrt s = 130A$ GeV. As expected the kaon radii fall 
between those for the proton and pion. The $K_T$ dependence for $K^- (K^+)$ radii 
is parameterized as $R_{\rm inv} = A/\sqrt{m_T}$, with $A=4.45 (4.74)$ fm GeV$^{1/2}$. 
The estimated kaon radii are consistent with that of the preliminary three-dimensional
measurements \cite{Eno} at the highest RHIC energy of $\sqrt s = 200A$ GeV.
We note that the $m_T$ dependence of the HBT radii $R_{\rm inv} = A/\sqrt{m_T}$ 
for these hadrons at RHIC are not consistent with one scaling coefficient as observed
at the SPS energy \cite{Murray}; the fitted values here are $A=3.81, 4.45, 5.32$
fm GeV$^{1/2}$ for $\pi^-,K^-,\overline p$.
The measured single particle $m_T$ spectra for $K^-$ and $K^+$ \cite{kaons} are
fitted with an exponential function $EdN/d^3p = dN/dy \exp[-(m_T-m_K)/T]/(2\pi
T(m_K + T))$ that results in a rapidity density of $dN/dy = 38.2 (42.1)$ for
$K^- (K^+)$ using a slope parameter of $T = 280$ MeV. Inserting the spectra
along with the the parameterization of $R_{\rm inv}$ into Eq. (\ref{eq:dsdy})
yields $dS/dy|_{K^-} = 256$ and $dS/dy|_{K^+} = 286$. We assume that the
neutral (anti)kaons carry the same entropy as the corresponding charged (anti)kaons.

The measured \cite{lambda} spectra for $\overline\Lambda$ for the 11\% most
central collisions at midrapidity has feed-down from the electromagnetic decay
of the $\overline\Sigma^0$ and the weak decays of the $\overline\Xi^+$,
$\overline\Xi^0$, and $\overline\Omega^+$. The feed-down from the charged
$\overline\Xi$ and $\overline\Omega$ states is corrected by using the
exponential parametrization of the the measured spectra for $\overline\Xi^+$
and $\overline\Omega^+$ of Ref. \cite{casom}. The corrected $\overline\Lambda$
spectra is then fitted with an exponential function resulting in $dN/dy = 7.4$
and slope parameter $T = 390$ MeV. The entropy rapidity density for
the antihyperons estimated by assuming that they have the same source
radii as the antiprotons are presented in Table I. To estimate
$dS/dy$ for $\overline\Sigma^\pm$, we note that the particle density in the
statistical model in the Boltzmann approximation is $N/V = g m^2 T K_2(m/T)
\exp(\mu/T)/(2\pi^2)$, where the notations are as usual. Assuming isospin
symmetry $\overline\Sigma^+ \approx \overline\Sigma^0
\approx \overline\Sigma^-$, and same temperature $T_{\overline\Lambda} \approx
T_{\overline\Sigma} = 173$ MeV, the thermodynamical model gives the ratio
$\overline\Lambda /\overline\Sigma^{\pm,0} = 1.47$ and thus
$\overline\Sigma^\pm = (\overline\Lambda + \overline\Sigma^0)/2.94$.  The
$dS/dy$ extracted under this assumption for $\overline\Sigma^\pm$ is listed in
Table I.

To extract the entropy density for baryons, we employ the measured ratio for
the yields $\overline p/p = 0.72$, $\overline\Lambda/\Lambda = 0.71$,
$\overline\Xi^+/\Xi^- = 0.83$, $\overline\Omega^+/\Omega^- = 0.95$
\cite{ratio}, and assume that all (anti)baryons have the same source radii as
antiprotons. Furthermore, the ratios are found to be independent over a large
range a $p_T$ suggesting that the transverse momentum distribution and thus the
slope temperatures of the baryons and their antibaryons can be taken to be
identical. The $dS/dy$ for the baryons thus estimated are shown in Table I.

The values of $R_{\rm inv}$ shown in Fig. \ref{fig:hbt} (top panel) may be
used in conjunction with the corresponding spectra of the hadrons to calculate
at a given $p_T$ the average phase space density at freeze-out 
$\langle f(p_T)\rangle = {\cal F}/2\sqrt 2$, where $\cal F$ is given by
Eq. (\ref{eq:dsdy}). This is displayed in Fig. \ref{fig:hbt} (bottom panel) where
the pion $\langle f\rangle$ at RHIC is remarkably higher than the values extracted 
at SPS \cite{na44,ferenc} and AGS \cite{e877} energies.

Under the assumption of isospin symmetry the total entropy rapidity density for
the hadron-gas at freeze-out for central collisions at RHIC is 4451.  Despite
the large uncertainty in source sizes, this number is surprisingly robust since
the radii only enter logarithmically. A 30\% change of $R_{\rm inv}^3$ changes
the entropy by 0.26 units per particle. For baryons, which have approximately
10 units of entropy per particle, the overall entropy would change by only a
few percent. Another source of error comes from estimating the contribution of
the $\eta$ and $\eta'$. However, since an $\eta$ carries two to three times
more entropy than a pion, and since the $\eta$ decays into two or three pions,
the net entropy is rather insensitive to the fraction of pions assigned to
$\eta$s. The same would be true if one were to explicitly include the entropy
of the $\phi$ meson while correspondingly reducing the kaon yields. Since the
yields are not measured to better than 10\% accuracy, the systematic error in
estimating the final-state entropy is estimated $\sim 10\%$.

We now estimate the entropy per rapidity density generated in the QCD partonic
matter in the early stage at RHIC. Assuming boost-invariant longitudinal
expansion, the energy density averaged over transverse coordinate is
\cite{Bjorken}
\begin{equation}\label{energy}
\overline{\epsilon_{\rm Bj}(\tau)} = \frac{dE_T}{dy} \frac{1}{\tau\pi R^2} ~.
\end{equation}
Using $dE_T/dy \approx 542$ GeV \cite{etp} and $\pi R^2= 120$ fm$^2$ for the
$\sim 11\%$ most central Au+Au collisions, one obtains
$\overline{\epsilon(\tau)} = 4.5$ GeV/fm$^3$ at a proper time $\tau = 1$
fm/c. Present data from lattice QCD simulation at $\mu_B=0$ shows a rapid
increase of $\epsilon/T^4$ by about an order of magnitude within a narrow
temperature range $\Delta T < 0.1T_c \sim 16$ MeV near the critical temperature
$T_c$ \cite{lattice}. This may indicate a first order transition, or nearly
first-order transition.  For three degenerate quark flavors, $n_f=3$, with a
mass $m/T=0.4$, transition to a plasma phase occurs at $T_c=154$ MeV at a
critical energy density $\epsilon/T_c^4 \simeq 6$.  Assuming the initial matter
at RHIC is already thermalized at 1 fm/c, lattice data indicates that
$\overline{\epsilon(\tau)} = 4.5$ GeV/fm$^3$ would correspond to an (average)
initial temperature of $T \approx 230$ MeV and pressure density $p
\approx 3T^4 = 1.1$ GeV/fm$^3$. The resulting entropy density for the 3 flavor
system is $s = (\epsilon +p)/T = 24.3$ fm$^{-3}$, and the average entropy
rapidity density at $\tau = 1$ fm/c is
\begin{equation}\label{lat3}
\frac{dS}{dy}\Big|_{{\rm QGP}; n_f=3} 
= \tau\int d^2x_T \: s  \: \simeq  2922 ~.
\end{equation}

%%%%%%%%%%%%%%%%%%%%%%%%%%%%%%%%%%%%%%
\begin{figure}[ht]
\centerline{\epsfig{file=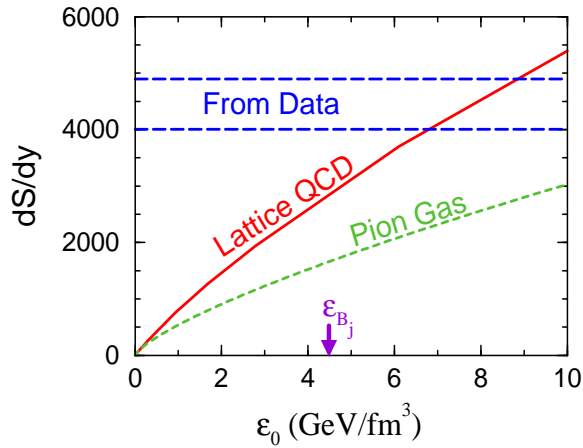,width=3.0in,height=2.4in,angle=0}}
\caption{The entropy per unit rapidity from lattice calculations (solid line)
is displayed as a function of the $\epsilon_0$, the energy density at $\tau=1$
fm/c. The horizontal band shows the final-state entropy extracted from
experiment. Hydrodynamic models typically have energy densities nearing 10
GeV/fm$^3$, which would push the limit for entropy from the final state. }
\label{fig:dsdy_vs_epsilon0}
\end{figure}
%%%%%%%%%%%%%%%%%%%%%%%%%%%%%%%%%%%%%%

The Bjorken estimate of the energy density only provides a lower bound since it
ignores the energy of the local longitudinal (along the beam axis) motion and
neglects the longitudinal work caused by the expansion of the plasma. The
inclusion of local longitudinal motion would increase the estimate by a factor
of $4/\pi$ if the emitted particles were massless and if there was no
transverse collective flow. Accounting for transverse flow and masses should
result into the increase being in the 10-15\% range. For massless particles
undergoing a purely one-dimensional Bjorken expansion, the energy density would
fall as $\tau^{-4/3}$, and $dE_T/dy$ would change by a factor of
$\tau^{-1/3}$. This would be a factor of 2 for times between 1 and 8 fm/c. The
loss of $E_T$ from longitudinal work is proportional to the $\pi R^2\int P
d\tau$ in hydrodynamic models, or $\pi R^2 T_{zz} d\tau$ if the stress tensor
is non-isotropic, i.e., viscous effects are included. The effects of viscosity
have been studied within the context of cascade models. Whereas,
one-dimensional hydrodynamic models might show that $dE_T/dy$ is higher than
that at breakup, the inclusion of viscosity reduces that increase to the
50\% range \cite{pang,cheng}. After including the effects of
longitudinal work and longitudinal motion, estimates of the initial average energy
density might be in the range of $7\pm 2$ GeV/fm$^3$, depending on the equation
of state and the viscosity. 

Of course, the initial entropy is less than the final state entropy. One
expects the entropy to increase during the reaction by the order of 10\% from
viscous effects or shocks \cite{seibertshocks}. If one were to lower the
final-state estimate of $dS/dy$ by 10\% it would come close to the value of
$dS/dy$ predicted by the lattice gauge calculation. A lack of complete
thermalization of the initial state would also point to a lack of entropy at
early times. If the initial energy was stored in classical gluon fields
\cite{cgc}, the entropy would be negligible. An underpopulation of quark
degrees of freedom would also lower the entropy. Since flavor and color charges
cannot sample the entire volume, finite-size effects can also reduce the
initial entropy, but only by a few percent \cite{joerg,elze}. Since the amount
of entropy produced during the expansion is unknown, it is impossible to rule
out that the initial state might have been of lower entropy.

Thus, our estimate of the final-state entropy, $dS/dy=4451$, does not seem
inconsistent with expectations from lattice calculations from analyzing
Fig. \ref{fig:dsdy_vs_epsilon0}. It therefore seems puzzling that hydrodynamic
calculations based on lattice-inspired equations of state matched spectra and
failed to fit HBT sizes \cite{Soff,teaneyhydro,heinzkolb,kolbheinz}. An
overestimate of a factor of two in $R_{\rm inv}^3$ would suggest an
overestimate of the entropy of the order of $\ln(2)$ per pion, or $\sim 400$
units per rapidity. However, it appears from the data that other degrees of
freedom, especially the baryonic degrees of freedom, compensate for the lack of
entropy in the pions. Although the pions comprise two thirds of the final-state
particles, they provide less than half the final-state entropy. The shift of
entropy from pions to the (anti)baryon sector was discussed by Rapp and Shuryak
\cite{rappshuryak} as a natural consequence of an increased phase space density
of pions. This follows from the reaction $p\overline p \leftrightarrow N\pi$. Since
pion number population is difficult to maintain, the overpopulation of pions
was expected in an isentropic expansion
\cite{greiner}. An additional factor in increasing the
overpopulation of pions would be a lowering of hadronic masses associated with
chiral symmetry restoration \cite{brownrho}. A lowering of hadron masses
would lead to an increase in the population of heavier hadrons, many of which
would decay back to multiple pions after the normal QCD vacuum is restored
\cite{haglinpratt}. 

It should be stressed that the small pionic sources extracted from HBT remain a
puzzle. Since the overall entropy of the final state is consistent with the
lattice gauge theory, the possibility remains open that a quark-gluon plasma
is indeed created at RHIC but that the dynamics of the expansion and decoupling
somehow differ from current hydrodynamic descriptions \cite{Hirano,KolbRapp}.

To illustrate the sensitivity of the entropy to the number of degrees of
freedom, Fig. \ref{fig:dsdy_vs_epsilon0} also displays $dS/dy$ for a pion
gas. Since the pion gas has only three effective degrees of freedom, it has a
much higher temperature for a given energy density than does the lattice
equation of state. From Eq. \ref{eq:DeltaS}, it therefore has a lower entropy
when compared at the same energy density. At high temperatures, a hadron gas
would effectively incorporate more degrees of freedom, and for energy densities
near or above 10 GeV/fm$^3$ would have more effective degrees of freedom than a
plasma due to the extremely large number of baryonic resonances with masses
between 1.0 and 2.0 GeV. However, since the volume of a baryon is in the
neighborhood of 2 fm$^3$, energy densities above 0.5 GeV/fm$^3$ preclude a
serious description in terms of a non-interacting hadron gas.

In conclusion, we have calculated the final-state entropy from $130A$ GeV
Au+Au collisions at RHIC to be $dS/dy=4451$. We estimate that this number,
which is based solely on measured spectra and extracted source-size radii of
the principal final-state constituents, to have a systematic uncertainty of
10\%. This number is not out of line with expectations from lattice gauge
theory.  Improved experimental measurements of spectra and source size might
reduce the systematic uncertainties to the 5\% level, but it is the uncertainty
in the reaction dynamics between the initial and final stages of the collision
that precludes our making a more definitive conclusion about the initial state
entropy. Although our number represents an important upper bound on the initial
state entropy, the amount of entropy generated throughout the reaction and the
initial energy density remain objects of conjecture. These issues will be
settled largely by other observables. For instance, elliptic flow measurements
are sensitive to the viscosity of the initial stages \cite{molnar} and HBT
measurements provide important three-dimensional constraints about the
dynamics. 

\bigskip

\begin{acknowledgments} This work was supported by the National Science 
Foundation under Grant No. PHY-0070818, and by the U.S. Dept. of Energy, 
Grant  DE-FG02-03ER41259.
\end{acknowledgments}

\end{document}